\documentclass[pra,aps,twocolumn,amssymb,showpacs]{revtex4}
\usepackage{graphicx}

\usepackage{epsfig}
\usepackage{graphicx}
\usepackage{dcolumn}
\usepackage{bm}

\begin{document}

\title{       Stripes and Pairing in High Temperature Superconductors
\footnote{\it This article is dedicated to Professor Karol I. Wysoki\'nski
              on the occasion of his 60th birthday.}
}

\author{ Andrzej M. Ole\'s }

\affiliation{Marian Smoluchowski Institute of Physics, Jagellonian
             University, Reymonta 4, PL--30059 Krak\'ow, Poland\\
             Max--Planck--Institut f\"ur Festk\"orperforschung,
             Heisenbergstrasse 1, D--70569 Stuttgart, Germany}

\date{30 November 2011}

\begin{abstract}
We review briefly several approaches used to investigate the stability
of stripe phases in high temperature superconductors, where charge
inhomogeneities arise from competing kinetic and magnetic energies.
The mechanism of stripe formation,
their consequences for the normal state and enhancement of pairing
interaction triggered by charge inhomogeneities are briefly summarized.
Finally, we demonstrate that orbital degeneracy ($i$) leads to a more
subtle mechanism of stripe formation, and ($ii$) plays an important
role and determines the symmetry of the superconducting state in pnictides.

{\it Published in: Acta Phys. Polon. A {\bf 121}, 752 (2012).}
\end{abstract}


\maketitle

\subsection{Models for CuO$_2$ planes}
\label{sec:intro}
\vskip -.3cm

There are several round anniversaries this year related to the research
on superconductivity --- not only 100 years since the discovery of this
phenomenon in mercury and 25 years since the discovery of high
temperature superconductors (HTSC). This latter discovery triggered
increased interest in the properties of strongly correlated electrons
on a two-dimensional (2D) square lattice. The simplest model to
describe the behavior of correlated electrons in HTSC is the Hubbard
model, or the effective $t$-$J$ model derived from it 35 years ago
\cite{Cha77}. This model was ready at the moment when the HTSC were
discovered and could be used to investigate their behavior
\cite{Oga08}.

At present we know that the Hubbard (or $t$-$J$) model provides the
simplest description of the electronic states in CuO$_2$ planes,
being a common structural element of HTSC. The minimal realistic
model for CuO$_2$ planes is the so-called three-band $d$-$p$
(charge-transfer) model, introduced simultaneously by a few
theoretical groups \cite{Eme87,Ole87,Var87} --- it contains
hybridized Cu($3d_{x2-y2}$) and O($2p_{\sigma}$) orbitals, with
Coulomb on-site interactions $U_d$ and $U_p$ between two
electrons with opposite spins, and the intersite Coulomb repulsion
$U_{dp}$. At this early stage of the theory of
superconductivity in HTSC it was important to establish how strongly
the $3d$ electrons are correlated. This was investigated using a
variational local ansatz to the three-band model \cite{Ole87}.
This approach was also employed at that time to establish universal
features concerning the correlation strength for $\sigma$, $\pi$ and
aromatic molecular bonds \cite{Ole86}. One finds that all these bonds
are weaker correlated than hybridized $d$-$p$ electrons, both in
CuO$_2$ planes \cite{Ole87} and in CuO$_3$ chains \cite{Ole91} found
in YBa$_2$Cu$_3$O$_7$ superconductor. This implies that charge
fluctuations are almost entirely suppressed and only certain
delocalization of holes centered at Cu sites over the neighboring
O($2p_{\sigma}$) orbitals takes place due to $d$-$p$ hybridization
$t_{pd}$ \cite{Ole87}. On the other hand, the effective superexchange
interaction along the Cu--O--Cu bonds between the $S=1/2$ spins at Cu ions
leads to an antiferromagnetic (AF) order \cite{Zaa88}. This result served as a
justification to use the effective Hubbard model instead of the
more complete three-band model.

In HTSC it is more convenient to introduce the hole notation as the
reference state in La$_2$CuO$_4$ contains one hole at each Cu$^{2+}$
ion in the $d^9$ configuration. Doping in La$_{2-x}$Sr$_x$CuO$_4$ or
in YBa$_2$Cu$_3$O$_{6+x}$ (where the CuO$_3$ chains gradually form)
generates holes in oxygen orbitals within CuO$_2$ planes and leads to
formation of local Zhang-Rice singlets \cite{Zha88}. This concept is
crucial as a doped hole occupies not a Cu($d_{x^2-y^2}$) orbital but
a linear combination of $p_{\sigma}$ orbitals around a hole with
$x^2-y^2$ symmetry and forms a singlet together with the hole at the Cu ion.
Therefore, adding a hole into a CuO$_2$ plane may be viewed as a
removal of one Cu spin in the effective Hubbard model that describes
the Mott insulator,
\begin{equation}
\label{Hub}
H=-\sum_{ij,\sigma}t_{ij}
a^\dagger_{i\sigma}a^{}_{j\sigma}
+U\sum_in_{i\uparrow}n_{i\downarrow}\,,
\end{equation}
where $a^\dagger_{i\sigma}$ is a hole creation operator in an
$|i\sigma\rangle$ state and $n_{i\sigma}=a^\dagger_{i\sigma}a^{}_{i\sigma}$.
Here $t_{ij}\equiv t$ or $t_{ij}\equiv t'$ are hopping elements between
the nearest neighbor (NN) or next-nearest neighbor (NNN) Cu ions, and
$U$ is an effective on-site
Coulomb parameter. The latter stands for the charge-transfer
gap $\Delta$ in the electronic structure of the three-band $d$-$p$ model
and is much lower than $U_d\simeq 10$ eV. The parameters in
Eq. (1) can be derived from the electronic structure calculations
\cite{Gra92} (see also \cite{Ole10}), and
one finds $t\equiv t_{pd}^2/\Delta\simeq 0.4$ eV and $U\simeq 4$ eV.

The three-band model with the Zhang-Rice singlets and the above Hubbard
model (\ref{Hub}) are two independent routes toward the effective
$t$-$J$ model, derived using perturbation theory in Cracow 35 years ago
(and published one year later \cite{Cha77}) --- another
round anniversary --- this model describes the electronic states in
cuprates \cite{Oga08}. A properly chosen canonical transformation leads
from the full Hilbert space to an effective low-energy Hamiltonian
acting in the restricted space. The model consists of the kinetic energy
$\propto t$ (here we assume $t'=0$) and the AF superexchange
$\propto J\equiv 4t^2/U$ between $S=1/2$ spins:
\begin{equation}
\label{tJ}
{\cal H}_{t-J}\!=\!-\sum_{ij,\sigma}t_{ij}
{\tilde a}^\dagger_{i\sigma}{\tilde a}^{}_{j\sigma}
+J\sum_{\langle ij\rangle}\left({\bf S}_{i}\!\cdot\!{\bf S}_{j}
-\frac{1}{4}{\tilde n}_i{\tilde n}_j\right)\,.
\end{equation}
The operators ${\tilde a}^\dagger_{i\sigma}=a^\dagger_{i\sigma}(1-
n^{}_{i\bar{\sigma}})$ ($\bar{\sigma}=-\sigma$) are projected fermion
operators and act in the restricted space. The above $t$-$J$ model
(or $t$-$t'$-$J$ model when $t'\neq 0$) follows also directly from
the three-band $d$-$p$ model --- the superexchange in cuprates
includes both the Anderson and charge-transfer excitations \cite{Zaa88}
and stabilizes 2D AF order at $x=0$ \cite{Bir98}. The value of
$J\simeq 0.13$ eV is either deduced from the magnetic experiments, or
derived from the parameters of the charge-transfer model \cite{Gra92}.

The first intriguing question concerning hole doping is whether a
hole doped to the AF state may propagate coherently. Naively one might
argue that a hole creates defects on its way, so it would need to make
a hopping along a closed loop to annihilate these defects and to
propagate in the square lattice \cite{Tru88}. Actually,
this is the only possible process to delocalize the hole in the Ising
model, see also Sec. 4. But when a hole is doped into a Heisenberg
antiferromagnet the quantum fluctuations of the AF background repair
the defects created by the hole, resulting in the propagation of a hole
in form of a quasiparticle (QP) with
dispersion on the energy scale of $J$ \cite{Mar91}. This concept was
confirmed by angle resolved photoemission (ARPES) experiments in
cuprates \cite{Dam03}. Detailed comparison between the experimental
data of ARPES experiments and the outcome of the theoretical
calculations performed using the self-consistent Born approximation
(SCBA) \cite{Mar91} were presented by several groups. One of the
highlights is the theoretical explanation of high quality ARPES data
obtained for Sr$_2$CuO$_2$Cl$_2$ which are reproduced by the SCBA
calculations performed for $t'=-0.3t$ \cite{Bal95}. This demonstrates
that the $t$-$t'$-$J$ model is the right effective model for the HTSC.

The remaining of this paper is organized as follows. We analyze the
origin and physical properties of stripe phases in section 2, and show
in section 3 that they may enhance the pairing interaction. Stripe
phases may also form in systems with alternating orbital order but
their origin is more subtle as shown in section 4. There we also
point out that Hund's exchange and orbital physics play an important
role in pnictides. Short summary is presented in section 5. Original
graphical material on the stripe phases may be found in the cited
literature.

\subsection{Microscopic origin of stripe phases}
\label{sec:stripes}
\vskip -.3cm

Doping of CuO$_2$ planes weakens AF correlations which become
short-range but survive up to the overdoped regime ($x\simeq 0.2$)
\cite{Bir98}. While AF and superconducting (SC) states exist in distinct regimes, there
are several possibilities how the phase diagram could look like when
doping increases \cite{Dag05}. In fact, the doped holes
self-organize in form of phases with charge and magnetization density
modulation, called {\it stripe phases\/} \cite{Voj09}. Such structures,
with the charge density varying twice faster than the magnetization
density in real space \cite{Zaa89,Zaa96}, were first discovered in the
theory as an instability of doped antiferromagnets.
Only a few years later their existence in cuprates
was confirmed by neutron experiments \cite{Tra96}.

The first question concerning stripe phases is whether they are
as solitonic defects in the AF structure, i.e., separating different
AF domains, or they form as polarons in a single AF domain. Although
naively one could argue that usually polarons optimize better the
kinetic energy, the answer is more subtle. The simplest stability
estimate is obtained by considering a three-site cluster filled by two
electrons and centered at the domain wall \cite{Ole00}.
Due to self-organization
at large $U\gg t$, the electrons are confined within the cluster and
their configurations with either identical or opposite spins correspond
to: ($i$) polaronic, and ($ii$) solitonic unit, with energies $E_P$ and
$E_S$. One finds $E_P=-\sqrt{2}t$ and $E_S=E_P-4t^2/U$, i.e., the
solitonic energy is lower by the superexchange energy $J$ due to the
three-site effective hopping processes \cite{Ole00}. This
simple argument explains the experimental finding \cite{Tra96} that
charge walls are nonmagnetic and separate domains with different phase
of the AF order parameter.

First quantitative results for the stripe phases were obtained within
the Hartree-Fock (HF) approximation \cite{Zaa89,Zaa96}, and then
refined using variational wave functions \cite{Gor99} and rotationally
invariant version of the slave-boson approach formulated as renomalized
mean field theory (RMFT) with Gutzwiller renormalization
\cite{Rac06}. The HF studies gave remarkably robust insulating vertical
(or horizontal) (01) stripe structures with the observed filling of one
doped hole per two domain wall sites \cite{Tra96}, and with rather large
charge density modulation between the centers of AF domains and the
domain walls \cite{Zaa96}. The smallest charge unit cell found contains
four atoms which explains that the stripes melt in the overdoped regime
$x>0.19$. Evolution of the electronic structure described within the HF
studies shows systematic changes of the Fermi surface, depending on the
ratio of the charge incommensurability and hole doping $x$ \cite{Ich99}.

In the HF approach the stripes are stabilized by certain
additional (spin or charge) density modulation along the domain walls
\cite{Zaa96}, so one expects major changes when electron correlations
are included. Indeed, the modulation of charge and magnetization
density was reduced from the HF values within
the RMFT for the 2D Hubbard model \cite{Rac06}. This approach allowed
one to treat strong electron correlations in stripe phases with
large unit cells relevant in the low doping regime, and gave stable
stripes in the thermodynamic limit. It also helped to resolve
the longstanding controversy concerning the role played by the kinetic
energy in the stripe phases. While the transverse hopping across the
domain walls yields the largest kinetic energy gain in case of
insulating stripes with one hole per site, the holes propagating
{\it along\/} the domain walls stabilize instead metallic vertical
(01) stripes, with one hole per two sites in the cuprates.

Although the long-range Coulomb interaction might help to stabilize
the stripe order further, the great success of the real-space
dynamical mean field theory (DMFT) studies for the 2D Hubbard model
was the proof that the correct treatment of the on-site interaction
alone suffices to stabilize experimentally observed metallic stripes
in a rather broad doping range $0.03<x<0.2$ \cite{Fle00}. These
calculations reproduced also the observed crossover from diagonal (11)
to vertical (01) site-centered stripes at doping $x\simeq 0.05$
\cite{Yam98}. In addition, also the doping dependence of the size of
magnetic domains and the shift of the chemical potential,
$\Delta\mu\propto -x^2$, were found to be in quantitative agreement
with the experimental results for La$_{2-x}$Sr$_x$CuO$_4$. In this way
the paradigm of insulating stripe phases was abolished --- the chemical
potential was decreasing with doping within the metallic phase.

The spectral functions obtained within the DMFT \cite{Fle00} show
a coexistence of the incoherent states in the lower Hubbard band
and coherent QP states close to the Fermi energy. The main features of
the spectra are: a flat part of the QP band near the $X=(\pi,0)$ point,
and gaps for charge excitations at the $Y=(0,\pi)$ and
$S=(\pi/2,\pi/2)$ points in the low-doping regime $x<1/8$. These gaps
are gradually filled and close up under increasing doping, in
agreement with the experimental ARPES data for La$_{2-x}$Sr$_x$CuO$_4$
\cite{Ino00}. In a range of low temperature
the obtained spectra have a distinct QP peak at the $X$
point, present just below the Fermi energy $\mu$, a charge gap,
and a distinct QP at the $S$ point \cite{Fle00}. At increasing
temperature the spectral function $A({\bf k},\omega)$ gradually
changes, indicating the melting of stripe order irrespective of the
choice of the NNN hopping $t'$ \cite{Rac10}.

These calculations demonstrated \cite{Fle00} the importance of
dynamical correlations which strongly screen the local potentials
resulting from the on-site Coulomb interaction $U$ and lead thus to
drastic changes in the distribution of spectral weight with respect
to the HF picture. It was also shown \cite{Fle00} that the melting of
stripe order is influenced by the NNN hopping element $t'$, which
also explains the observed difference
in the spectral properties between Bi$_2$Sr$_2$CaCu$_2$O$_{8+y}$
\cite{She95} and La$_{2-x}$Sr$_x$CuO$_4$ \cite{Ino00}. At the same
time, $t'$ can tip the energy balance between the filled diagonal
and half-filled vertical stripes \cite{Rac06}, which might
explain a change in the spatial orientation of stripes observed in
the HTSC at doping $x\simeq 1/16$.

\subsection{Charge inhomogeneities and pairing}
\label{sec:pair}
\vskip -.3cm

After understanding the mechanism of stripe formation in cuprates,
a natural question to ask concerns their role in the phenomenon of
superconductivity. This subject is relatively new and currently under
investigation by several groups. While low temperature properties of
the SC state cannot be explained within the original resonating
valence-bond (RVB) framework \cite{Oga08,Gan05}, evidence accumulates
that SC correlations coexist in these systems with charge
inhomogeneities. Microscopic evidence of inhomogeneities given by
scanning tunnelling microscopy (STM) \cite{Koh07} motivated several
recent studies in the theory. The simplest mean-field Hamiltonian for
a singlet superconductor with disorder in pairing interaction gives
indeed a distribution of gap values in the local density of states
(DOS) found by solving self-consistently Bogoliubov-de Gennes
equations \cite{Non05}. The above disorder follows here from dopant
interstitial O atoms which modify the local electronic structure
\cite{He06}.

Pioneering work in this respect was presented by Ma\'ska {\it el
al.\/} \cite{Mas07} who provided a beautiful explanation of the
origin of the experimentally observed positive correlation between
the magnitude of the SC gap and positions of dopant oxygens in
Bi-based superconductors. Their approach demonstrated that charge
inhomogeneities, caused by spatial variation of the atomic levels,
are responsible for local enhancement of the pairing interaction
$J$ in the effective $t$-$J$ model. However, in the more complete
three-band model one finds that the sign of the correction to $J$
depends on the potentials on nearby sites induced by the dopant
impurity, suggesting important corrections beyond the one-band model
\cite{Roser}.

Coexisting charge modulation and unidirectional $d$-wave SC
domains were found using the $t$-$J$ model at $x=1/8$ doping
\cite{Rac07}. In this study half-filled charge domains separated
by four lattice spacings were obtained along one of the crystal
axes leading to modulated superconductivity with {\it
out-of-phase\/} $d$-wave order parameters in neighboring domains.
Both the RMFT and variational Monte Carlo calculations yield that
the unidirectionally modulated superconducting phases are energetically
remarkably close to the uniform RVB phase, so they could easily be
stabilized by impurities or other effects going beyond the $t$--$J$
model \cite{Rac08}. These studies were also extended to the pyrochlore
lattice, where the phase diagram includes superconductivity coexisting
with the underlying valence-bond solid order \cite{Rac09}. This
interesting work demonstrates that the interplay between electron
correlation and geometrical frustration can stabilize novel states
of matter exhibiting microscopic coexistence of superconductivity
and spin dimer order.

Exact diagonalization of finite clusters of size $N$ (with periodic
boundary conditions) is an ideal unbiased method to investigate pairing
in models of interacting electrons \cite{Bon89}. This method was used
recently to investigate pairing in inhomogeneous $4\times 4$ clusters
described by the Hubbard model (\ref{Hub}) with inequivalent hopping
integrals along two directions in a 2D plane, $t_a$ and $t_b<t_a$, in
two patterns: checkerboard and stripe phase \cite{Tsa08}.
The pair binding energy at doping $x=M/N$,
\begin{equation}
\label{delta}
\Delta_{\rm B}(x)=2E_0(M)-\{E_0(M+1)+E_0(M-1)\}\,.
\end{equation}
is defined by comparing the ground state energies $E_0(M)$ obtained for
systems with separated holes and a hole pair for electron number $M$
\cite{Bon89}. The pair-binding energy $\Delta_{\rm B}(x)>0$ is found in
both considered structures with purely repulsive interactions.
At $x=1/16$ it has the largest value in the intermediate regime of
$t_b\simeq 0.6t_a$ and $U\simeq 8t_a$ for the checkerboard lattice,
but also for the stripe lattice this energy has a maximum close to
$U=12t_a$ \cite{Tsa08}.

Further evidence that charge inhomogeneities enhance superconductivity
has accumulated from cluster dynamical studies for the Hubbard model
\cite{Oka08}. The electronic properties of multilayers
with combinations of underdoped and overdoped layers were investigated
using cluster DMFT superlattices. It has been found that the SC order
parameter is enhanced by the proximity of the strong pairing scale
originating from the underdoped layers and can even exceed the maximum
value in uniform systems. In quantum Monte Carlo simulations stripe-like
charge-density wave modulation was imposed by a periodic potential
with modulation strength $V_0$ and the pairing correlations and
critical temperature were determined from the Bethe-Salpeter
equation in the particle-particle channel \cite{Oka08}. The optimal
superconductivity is then obtained for a moderate modulation strength
due to a delicate balance between the modulation enhanced pairing and
suppression of the pure particle-particle excitations by a
modulation reduction of the QP weight.

Recently stripe phases were also investigated in the 2D $t$-$J$ model
by means of infinite projected entangled-pair states \cite{Cor11}. The
states with stripe order were found to have a lower variational energy
than uniform phases, and the stripes support $d$-wave pairing. For a
fixed unit-cell size the energy per hole is minimized for a hole
density of 1/2 hole per unit length of a stripe. These results support
earlier findings obtained within the RMFT \cite{Rac06} and the DMFT
\cite{Fle00}. In addition, one finds that the pairing amplitude
is largest at the atoms with enhanced hole density and suppressed
magnetization, and the mean pairing amplitude has a characteristic
maximum as a function of hole density per unit length of stripe
\cite{Cor11}.
Mean field pairing theory for the charge stripe phase of HTSC was just
formulated \cite{Lod11} --- the model describes the most prominent
properties of the stripe phase remarkably well and predicts a pair
density wave with spatial modulation of the pairing amplitude in the
striped structure.

Experimental evidence accumulates that stripe order coexists with
the SC order. In fact, as the most likely interpretation of the
complete collection of results obtained for magnetic
susceptibility, thermal conductivity, specific heat, resistivity
and thermopower for La$_{1.875}$Ba$_{0.125}$CuO$_4$, appearance of
unusual 2D SC correlations together with the onset of spin-stripe
correlations was suggested \cite{Tra08}. This suggestion seems to reflect the
real situation as confirmed by extensive studies of La$_{2-x}$Ba$_x$CuO$_4$
compounds in the broad range of doping, $0.095<x<0.155$
\cite{Huc11}. Stripe incommensurability increases here with $x$ in
a smooth way, unlike in the earlier data for
La$_{2-x}$Sr$_x$CuO$_4$ \cite{Yam98}. In the entire range of
doping $x$ the charge order appears at a higher temperature
$T_{\rm CO}$ than the onset of spin order at $T_{\rm SO}$. Truly
static spin order sets in below the charge order and coincides
with the first appearance of in-plane SC correlations at
temperatures significantly above the SC transition in the bulk.
They also presented an interesting phase diagram \cite{Huc11} and
argued that charge order is the dominant order that is compatible
with SC pairing but competes with SC phase coherence.

Local enhancement of pairing mechanism may occur even in absence of
stripe order, as long as charge inhomogeneities are induced by structure
charges. For instance, apical oxygens may affect the strength of the
pairing potential within CuO$_2$ layers \cite{Mor08}. In fact, the
superexchange is very sensitive to the covalency which changes depending
on the position of apical oxygens. This is due to high polarizability of
O$^{2-}$ anion which has the effect of screening and reduces the Coulomb
interaction $U$ \cite{Gas84}. Hole density inhomogeneities that exist in
HgBa$_2$CuO$_{4+\delta}$ and Bi$_2$Sr$_2$CaCu$_2$O$_{8+\delta}$ are
expected to play a similar role and enhance the pairing locally
\cite{Che11}.

\subsection{Orbital degeneracy and pnictides}
\label{sec:orbi}
\vskip -.3cm

In contrast to cuprates, iron-based superconductors contain
several partly filled bands and orbital degeneracy plays a role.
Local electron interactions are then described by degenerate Hubbard
model with two parameters: intraorbital Coulomb element $U$ and Hund's
exchange $J_H$ \cite{Ole83}. When the spin-orbital superexchange model
is derived for such a Mott insulator at $t\ll U$, the multiplet
structure of excited states decides about different spin-orbital terms
that depend on $J_H$ and determine both the magnetic properties and
the optical spectral weights \cite{Kha04}. For instance,
the superexchange in manganites explains the observed spin and orbital
order in LaMnO$_3$ \cite{Fei99}, while the relevant spin-orbital
$t$-$J$ model for $e_g$ electrons gives a correct description of the
metallic FM phase including the origin of spin excitations \cite{Fei05},
and was also used to investigate the evolution of magnetic order with
increasing doping in doped momolayer and bilayer manganites \cite{Dag07}.

Stripe phases appear also in doped transition metal oxides with active
orbital degrees of freedom and some examples were given in Refs.
\cite{Ole10,Dag05}. Here we address briefly only the simplest case of
stripe order in doped La$_{2-x}$Sr$_x$NiO$_4$ with AF order
found about the same time as in cuprates \cite{Tra95}. In contrast to
cuprates, however, the stripes in La$_{5/3}$Sr$_{1/3}$NiO$_4$ are
diagonal and contain one hole per unit cell \cite{Wak09}. This
difference follows from orbital degeneracy and realistic $e_g$ hopping
that does not conserve the orbital flavour \cite{RacNi}.
Electronic structure calculations suggest that a subtle interplay
between the charge and spin order and octahedral distortions is
essential for the formation of an insulating state \cite{Sch09}.
Indeed, both local electron correlations and the Jahn-Teller
interactions with lattice distortions play a role in reproducing
experimentally observed stripes at $x=1/3$ doping and the checkerboard
structure at $x=1/2$ \cite{Ros11}.

In systems with active $t_{2g}$ degrees of freedom and FM spin order,
alternating orbital order follows from Ising-like superexchange and
stable stripe phases are different. The orbital $t$-$J$ model that
follows in this case from the degenerate Hubbard model on
the square lattice \cite{Woh08} is applicable either to transition
metal oxides with active $t_{2g}$ orbitals filled by one electron
at each site (with $\{yz,zx\}$ doublet), or to cold-atom systems
with active $p$ orbitals. Although the Ising superexchange suggests
that the holes would be confined, the system may self-organize at
finite doping in a form of stripe phase \cite{Che00}. In cuprates
the presence of quantum spin fluctuations favors stripes in form of
ladders with dominating singlet correlations on the rungs \cite{Wro06}.
Here quantum spin fluctuations are absent and orbitals reorient into
ferro-orbitally ordered domain walls that allow for deconfined motion
of holes along them \cite{Wro10}. These solitonic stripes are more
stable than polaronic stripes, but the phase change of the
staggered order by $\pi$ plays a minor role in orbitally ordered
systems and both types of stripes might be expected at finite
temperature. The ferro-orbital order is induced by the
kinetic energy of doped holes and occurs here by the same
mechanism as in $e_g$ orbital systems \cite{Dag04}.

Coming to iron-based pnictides, several families of them exist (for
more details see a recent review \cite{Pag10}), and it is a challenge
for the theory to explain the origin of the SC pairing and its
symmetry. It was initially speculated that the pairing in these new
superconductors is related to that of the cuprates, but {\it de facto\/}
the situation is by far more complex. First, similar to cuprates, AF
interactions decide about magnetic order in pnictides in the vicinity
of SC states \cite{Neu11}. One finds a spin-density wave or a $C$-type
AF ($C$-AF) phase, sometimes incorrectly called "a stripe phase" (this
phase occurs here without doping). Second, the electronic structure of
pnictides was investigated in detail and it was shown that all $3d$
states of Fe ions are partly filled \cite{Lilia} and {\it a priori\/}
play an important role in the electronic instabilities. Electronic
structure calculations predict the magnetic order with large magnetic
moments but the observed moments are small $\sim 0.2\mu_{\rm B}$.
This problem can be resolved \cite{Lilia} by reducing the Stoner
parameter $I$ which contains two systematic errors when derived from
local density approximation (LDA) \cite{Sto90}. In fact, the LDA+DMFT
calculation demonstrate that \cite{Hau09}: ($i$) Hund's exchange $J_H$
is important and decides about the actual
values of the resistivity and specific heat in pnictides, and
($ii$) the magnetic moment in the $C$-AF phase is much reduced from
the LDA value.

It has been established by now that the superconductivity in layered
iron-based materials occurs by unconventional mechanism \cite{Boe08}.
Impurities play a similar role as in cuprates and locally modify
pairing conditions in pnictide superconductors \cite{Cie09}.
The structure of the Fermi surface is rich and several orbital states
could be involved in the pairing \cite{Lilia}. Therefore, next to AF
interactions, the orbital degeneracy plays here a very important role
and the microscopic models have to treat explicitly at least two
orbitals per site. Interorbital pairing interactions arise in such a
model and stabilize $s_{\pm}$-wave symmetry of the superconducting
order parameter \cite{Cie09}. However, the symmetry of the
superconducting state in pnictides is still controversial
and currently under investigation.

The two-orbital model was suggested shortly after
these systems were discovered as a generic model to investigate both
the magnetic order and the pairing \cite{Dag08}. This model includes
$t_{2g}$ partly filled orbitals $\{xz,yz\}$ and the electronic structure
consists of a wide band and a narrow band \cite{Rag08}. Keeping only
these orbitals is reasonable knowing that $xz$ and $yz$ orbitals provide the
largest contribution to the pnictides' Fermi surface \cite{Lilia}.
Superexchange interactions obtained in the spin-orbital model are here
frustrated as both nearest neighbor ($J_{\rm NN}$) and next-nearest
neighbor ($J_{\rm NNN}$) exchange is AF \cite{Si08}.

Recent Lanczos diagonalization studies of the $t$-$U$-$J$ model using a
small $\sqrt{8}\times\sqrt{8}$ cluster and Eq. (\ref{delta}) with $M=18$
to determine the pairing energy showed \cite{Nic11} that Hund's exchange
$J_H$ is crucial also here, together with the AF frustrated
superexchange. One finds that the $A_{1g}$ and $B_{2g}$ pairing
symmetries compete with each other in the realistic parameter regime.
Quasinodal $A_{1g}$ states are stabilized for physical values of $J_H/U$
and for sufficiently large $U$, in agreement with the RPA results
\cite{Gra09}, but $B_{1g}$ pairing symmetry is also energetically close
and could be stabilized by other weak interactions.
The two-orbital model is oversimplified but its results \cite{Nic11}
agree qualitatively with those obtained in the mean-field approximation
for the three-orbital model \cite{Dag10}. Here the pairing instabilities
also highlight the importance of Hund's exchange in pnictides.

\subsection{Summary and conclusions}
\label{sec:final}
\vskip -.3cm

Summarizing, stripe phases arise from a competition between the
superexchange energy of localized spins and the kinetic energy of doped
holes as a joint instability toward coexisting spin and charge
modulated order. Recent results suggest that the stripe phases may
coexist with superconductivity. In fact, the pairing interactions in
cuprates are enhanced by charge inhomogeneities that are generated by
doping. Further theoretical progress in the understanding of stripe
phases and their consequences for the pairing requires sophisticated
self-consistent cluster calculations for stripe phases.

In pnictide superconductors necessary ingredients of the theory are
orbital degeneracy and Hund's exchange. This makes the theoretical
studies of the pairing mechanism in pnictides more demanding but
also here the AF superexchange interactions are crucial for the pairing
mechanism.

\acknowledgments
\vskip -.3cm

It is a great pleasure to thank for so pleasant collaboration and
for numerous insightful discussions --- I thank in particular
Raymond Fr\'esard, Piotr Wr\'obel and Jan Zaanen,
as well as M. Daghofer, E. Dagotto, D. G\'ora, M. Fleck,
A.I. Lichtenstein, A. Moreo, D. Poilblanc, M.~Raczkowski,
K.~Ro\'sciszewski and G.A. Sawatzky.
We acknowledge support by the Polish National Science Center (NCN)
under Project No. N202 069639.

\end{document}